\documentclass[12pt]{article}
\usepackage{amsmath}
\usepackage{graphicx}
\usepackage{natbib}
\usepackage{url} 

\usepackage{amssymb}
\usepackage{lscape}
\usepackage[hyperindex]{hyperref}
\usepackage{epstopdf}
\usepackage{subfigure}
\usepackage{setspace}
\usepackage[normalem]{ulem}
\usepackage{verbatim}
\usepackage{color}
\usepackage{anyfontsize}
\usepackage{wrapfig}
\usepackage{sidecap}

\newtheorem{theorem}{Theorem}[section]
\newtheorem{lemma}[theorem]{Lemma}

\usepackage{natbib}

\newcommand{\blind}{0}

\begin{document}

\def\spacingset#1{\renewcommand{\baselinestretch}%
{#1}\small\normalsize} \spacingset{1}


\if0\blind
{
  \title{\bf Probabilistic Detection and Estimation of Conic Sections from Noisy Data}
  \author{Subharup Guha\thanks{
    The authors gratefully acknowledge partial support from the NSF grants DMS 1127914 given to the Statistical and Applied Mathematical Sciences Institute (SAMSI), which allowed the authors to collaborate on this project, and the project was initiated when the first author visited SAMSI for a year-long program. This work was also partially supported by  NSF grant
DMS-1854003  awarded to the first author. }\hspace{.2cm}\\
    Department of Biostatistics, University of Florida\\
    and \\
    Sujit K. Ghosh\\
    Department of Statistics, North Carolina State University}
  \maketitle
} \fi

\if1\blind
{
  \bigskip
  \bigskip
  \bigskip
  \begin{center}
    {\LARGE\bf Probabilistic Detection and Estimation of Conic Sections from Noisy Data}
\end{center}
  \medskip
} \fi

\bigskip
\begin{abstract}
Inferring unknown conic sections on the basis of noisy data is a challenging problem with applications in computer vision. A major limitation of the currently available methods for conic sections is that estimation methods rely on the underlying shape of the conics (being known to be ellipse, parabola or hyperbola). A general purpose Bayesian hierarchical model is proposed for conic sections and corresponding estimation method based on noisy data is shown to work even when the specific nature of the conic section is unknown. The model, thus, provides probabilistic detection of the underlying conic section and inference about the associated parameters of the conic section. Through extensive simulation studies where the true conics may not be known, the methodology is demonstrated to have practical and methodological advantages relative to many existing techniques. In addition, the proposed method provides probabilistic measures of uncertainty of the estimated parameters. Furthermore, we observe high fidelity to the true conics even in challenging situations, such as data arising from partial conics in arbitrarily rotated and non-standard form, and where a visual inspection is unable to correctly identify the type of conic section underlying the data.
\end{abstract}

\noindent%
\vfill

\newpage
\spacingset{1.5} 

\section{Introduction}\label{S:introduction}
In many practical scenarios in computer vision, camera calibration, and image recognition, among many others \citep[e.g.,][]{Ayache_1991, Forstner_1987, Beck_Arnold_1977, Ballard_Brown_1982, Szpak_etal_2012}, it is of interest to estimate curves or surfaces based on a set of noisy data. Specifically, consider a random sample of paired observations $\{(X_i, Y_i): i=1,2,\ldots,n\}$ which are assumed to arise from a joint distribution of  generic pair $(X, Y)$. It is known that $X=U+\epsilon_1$ and $Y=V+\epsilon_2$, where $(\epsilon_1, \epsilon_2)$ denotes bivariate independent noise or error, and that the support of the unobserved pair $(U, V)$  lies on a conic section.

\begin{figure}
\begin{center}
\vspace{- 1 in}
\begin{tabular}{cc}
\includegraphics[scale=0.4]{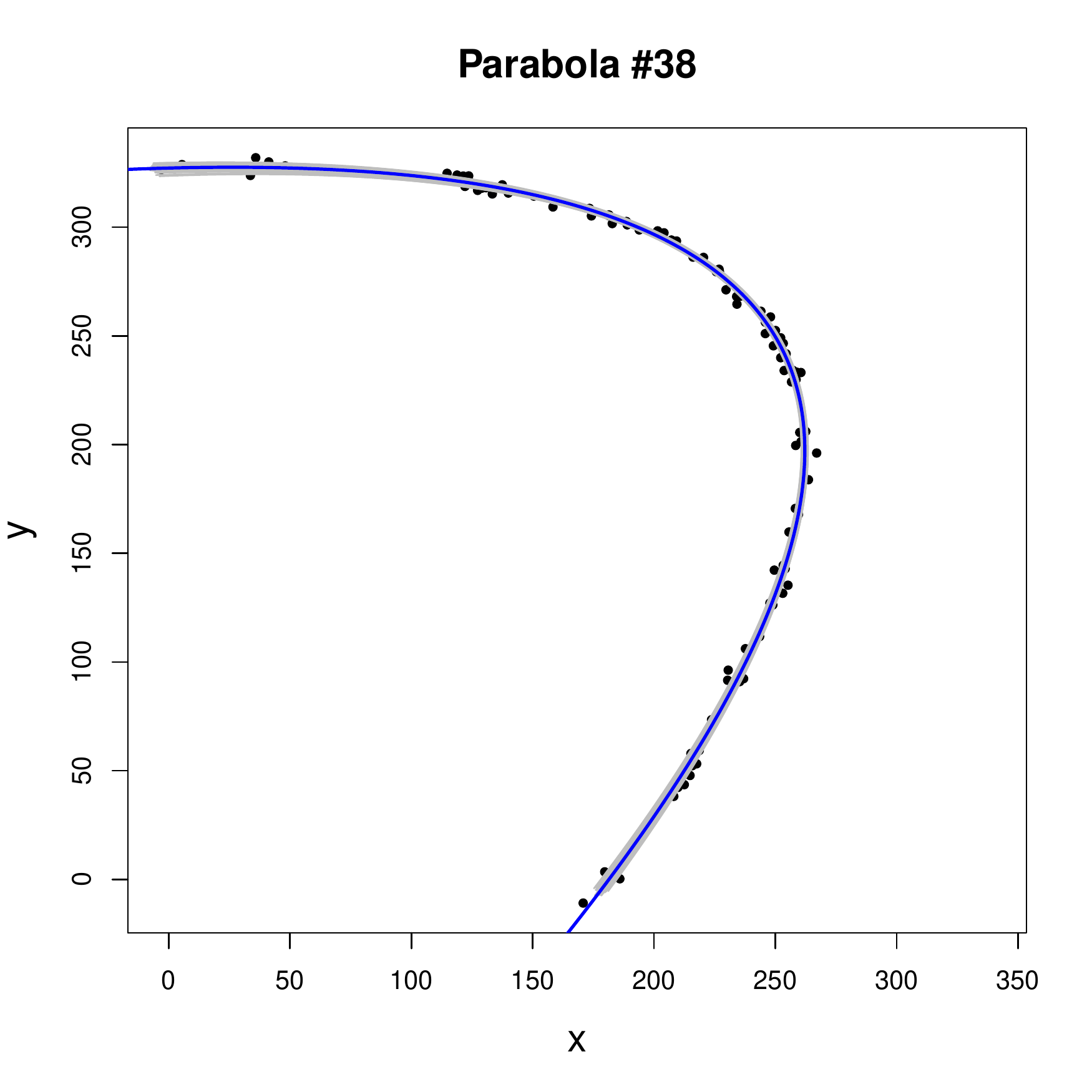} &
\includegraphics[scale=0.4]{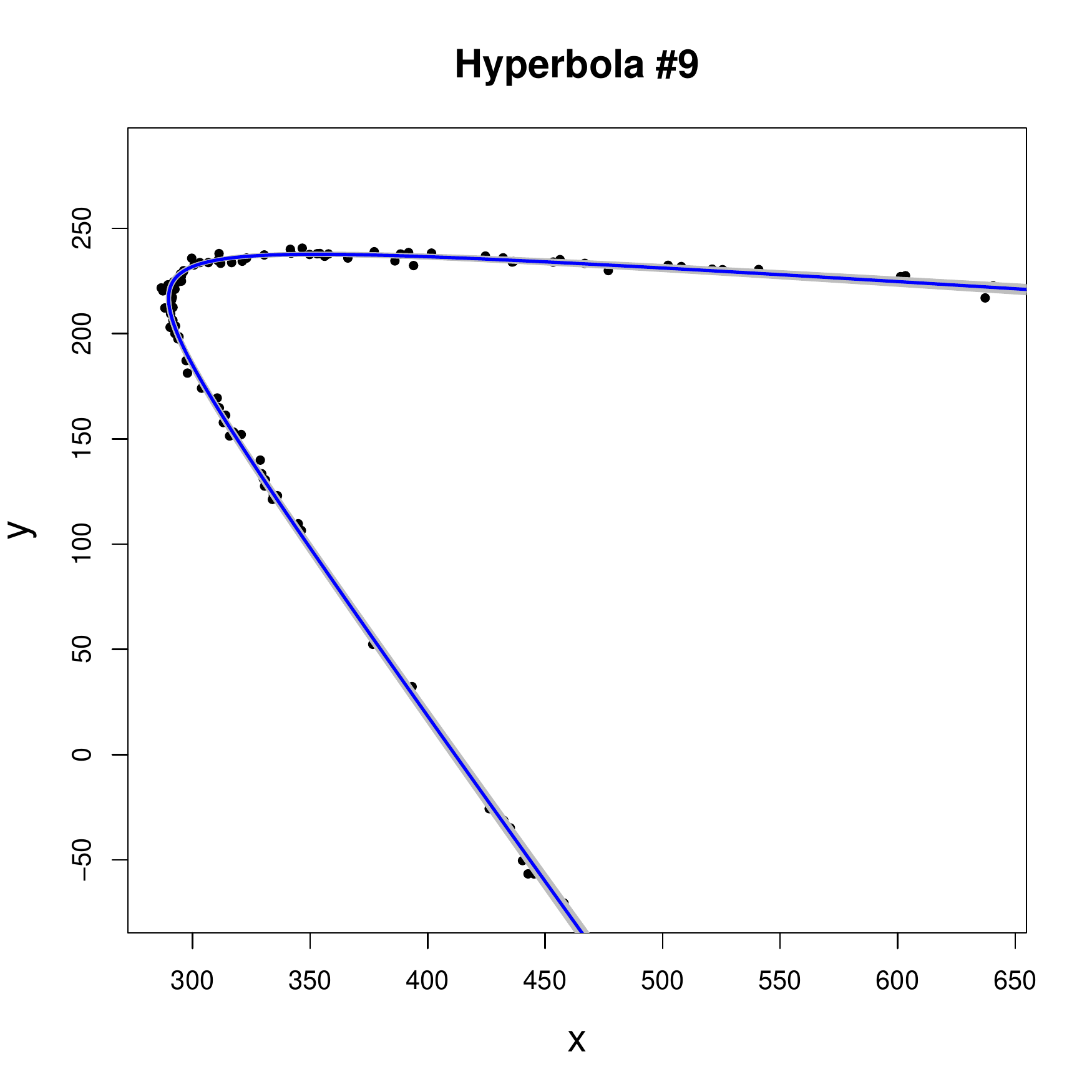} \\
\includegraphics[scale=0.4]{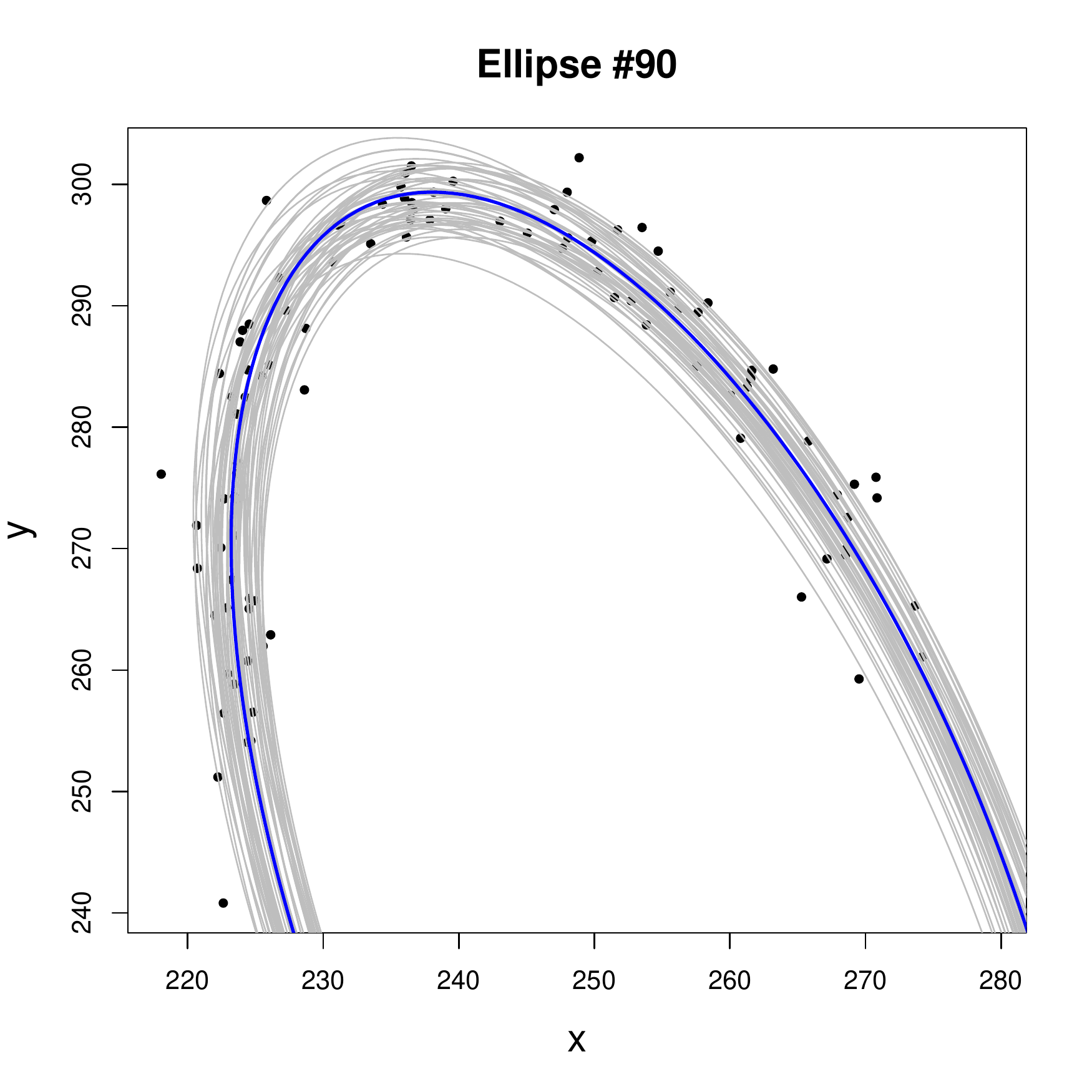} & .
\end{tabular}
\vspace{- .2 in}
\caption{A randomly chosen ellipse, hyperbola, and parabola from Simulation 2 of Section~\ref{S:simulation_1}. The blue bold lines indicate the estimated curves by the proposed method. The gray lines represent random samples drawn from conservative 95\% posterior credible intervals for the true curves. See the text for further explanation.}
\label{F:plot0}
\end{center}
\end{figure}

For instance, in Figure~\ref{F:plot0}, we illustrate three cases where the samples are drawn from three popular conic sections, parabola, hyperbola and ellipse. Without  first hand knowledge about the specific nature of the underlying support of the random pair $(U, V)$, it is visually difficult to guess the support of the pair of  random variables, even when we have first hand knowledge that the support belongs to a conic section. Moreover, the usual least squares based methods for a statistical model, i.e., $V=m(U)+\epsilon$, where $m(u)=E[V|U=u]$ denotes the regression function, are not readily applicable for these type of data and would often lead to erroneous inferences. This is primarily because a given value of $u$ does not correspond to a unique value of the underlying variable $V$. This motivates us to develop models and associated methodologies that would not only enable us to estimate the underlying parameters of a given conic section (e.g., being ellipse, parabola or hyperbola), but also provide  probabilistic detection of the hidden type of conic section.

In many applications, there are practical consequences of misclassifying the type of conics, e.g.\ incorrectly classifying a
parabola as a hyperbola.
A classic example is estimating the orbit of a smaller body (e.g.\ a planet or exoplanet) around a larger body (e.g.\ a star) which takes the shape of an ellipse, with the larger body being at one of the two focal points of the ellipse. This definitely has become of great significance in modern astronomy when discovering new exoplanets. The  website \url{https://exoplanets.nasa.gov/} has many recent applications. Another application is that of the correct identification and estimation of parabolas and hyperbolas in the field of optics. For example, car headlights are often in a parabolic shape because this causes a highly focused beam of light in front of the car and real-time estimation of its focus is of great importance in modern automated cars.
For many other interesting modern applications, refer to \cite{10.1007/978-3-319-95588-9_219}. 

For a quick glimpse of our proposed method, in  Figure~\ref{F:plot0}, we present the fitted conic section using the solid blue lines overlayed on a set of simulated data points generated from partial conics. The variance of the added bivariate Gaussian noise was the same in each dataset.  (However, the noise appears to be less for the parabola and hyperbola because they  are not compact, and their data have much greater ranges than the ellipse.)  The data were  analyzed assuming that we were unaware of the type of  true conic section. 
The estimation uncertainty is indicated by the gray curves, which represent samples drawn from conservative 95\%  posterior credible intervals for the conics, and correspond to multiplicity-adjusted, Bonferroni-type intervals for the univariate conics hyperparameters. The posterior probabilities of the detected types of conic sections for each of the three data sets are displayed in Table~\ref{T:simulation011}. We find that in all three examples, the probability of detecting the true type of hidden conic section is nearly perfect. This is particularly interesting given that data generated from a true parabola could  be easily misjudged to have come from a true ellipse, and vice versa.

\begin{table}[ht!]
\begin{center}
\renewcommand{\arraystretch}{1}
\begin{tabular}{ l  | c |c  | c}
\hline\hline
&\multicolumn{3}{c}{\textit{Posterior classification probabilities}}\\
\hline
 &\textbf{Ellipse}	&\textbf{Parabola}	&\textbf{Hyperbola} 	\\
\hline
\textbf{Parabola \#38} &0.005 (0.005) &0.995 (0.005) &0.000 \\\hline
\textbf{Hyperbola \#9}	&0.000 &0.000    &1.000\\\hline
\textbf{Ellipse \#90}	&1.000 &0.000    &0.000\\\hline
\hline
\end{tabular}
\end{center}
\caption{Estimated posterior probabilities of classification for the three  conic sections displayed in Figure~\ref{F:plot0}.
The estimated standard deviations of the non-degenerate probabilities are displayed in parentheses.}\label{T:simulation011}
\end{table}

In the vast literature on conics and in several engineering applications over the past several decades, many novel and computationally efficient methods have emerged that provide the `best' fit of the underlying conic section when it is known that underlying conic section is an ellipse (or circle), parabola or hyperbola \citep[e.g.,][]{Prasad_etal_2013, Ahn_etal_2001, Zhang_1997, Szpak_etal_2012}. However, to the best of our efforts, we were unable to find any methodologies that provide probabilistic inference about the underlying nature of the conic section based on noisy observations.

In essence, we provide an answer to the following motivating question: {\em Given a set of noisy observations, $(X_i, Y_i)$ where $i=1,\ldots,n$, and the fact that $(U, V)$ belongs to a conic section, can we estimate the probability that the support of $(U, V)$ is an ellipse, a parabola or a hyperbola?} Answers to such a question are of immense practical value in computer vision and image detection, as we will show that even when the observations arise from only a portion of the (unknown) underlying conic section, our proposed method is able to not only estimate the parameters of the  conic section but also provide an estimate of the probability of the detected conic type.

\smallskip

In order to define our models and associated methodologies, we first present some basic notions of  conic sections, which although  easily found in any elementary  mathematics textbook dealing with conic sections, helps us to fix the notation to be used throughout the rest of this paper.

A conic section is produced when a plane intersects with a right circular conical surface. The three types of conics that results are known as ellipses (which includes circles), parabolas, and hyperbolas. There are many different formulations for a conic section based on Cartesian coordinates (algebraic definition) or polar coordinates (based on focus-directrix definition). We start with the latter  as we find that formulation most useful for our purposes.

\subsection{Focus-directrix definition of conics using polar coordinates}\label{S:FD}

\begin{figure}[h!]
\begin{center}
\includegraphics[scale=0.75]{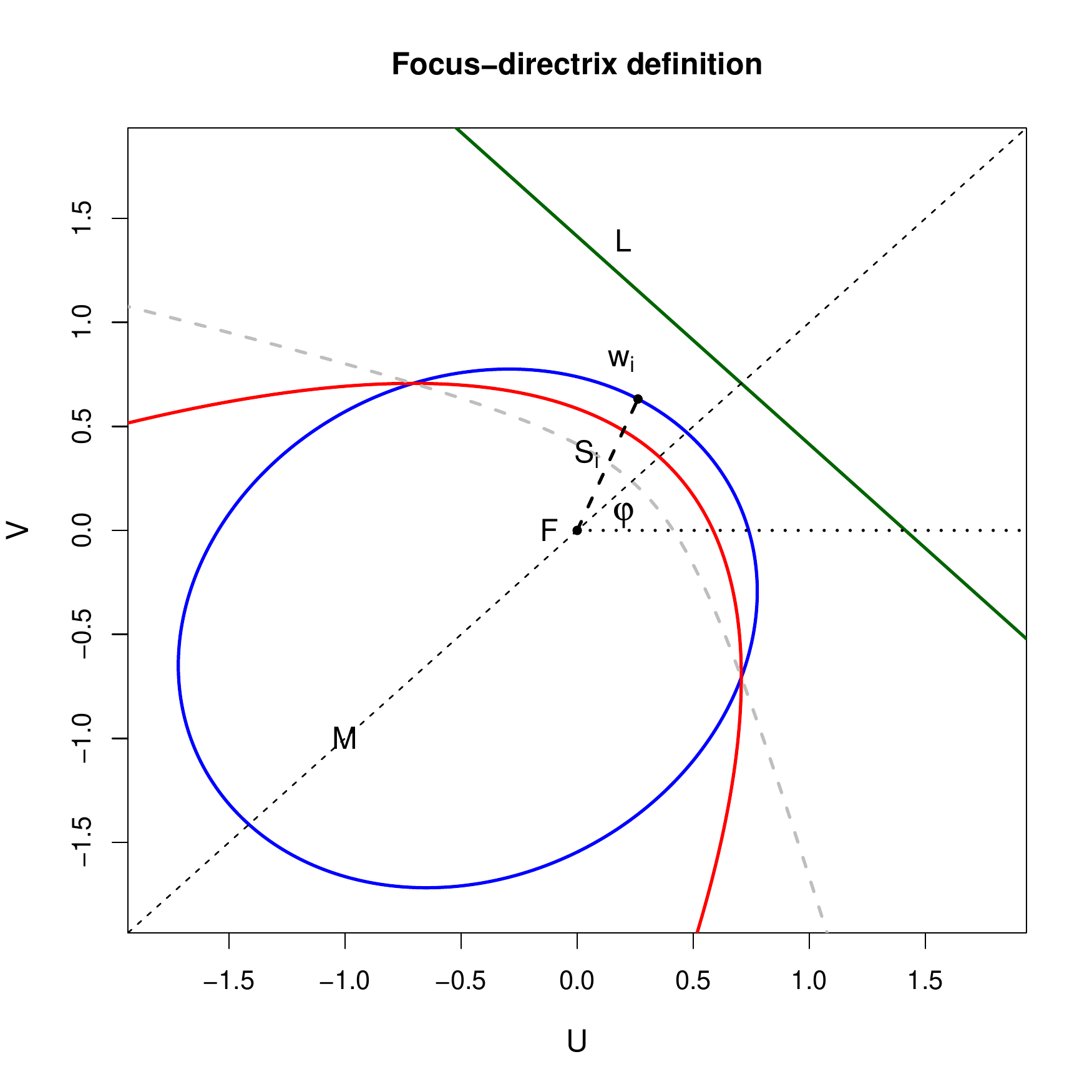}
\caption{An example with the three types of conic sections with a common focus $F$ located at the origin and a common directrix $L$ shown in green. An ellipse (blue, bold) with eccentricity $e=0.5$, parabola (red, bold), and hyperbola (gray, dashed) with  eccentricity $e=2$ have been plotted. Just a single branch of the parabola and hyperbola has been plotted; the other branch is obtained by reflecting on the directrix.
The  axis of symmetry, $M$, for the three conics is the dotted $45^\circ$ line. The common semi-latus rectum is $l=1$. The rotational angle is $\varphi=\pi/4$. A single point, $\boldsymbol{w}_i$, on the ellipse is shown with segment $S_i$ joining it to the focus. For this ellipse point, the length of the segment is $r_i = 0.684$ and the counter-clockwise angle from the axis of symmetry $M$ to the segment is $t_i=\pi/8$.}
\label{F:plot100}
\end{center}
\end{figure}

Among the different equivalent representations of conics, one of the most useful is the focus-directrix approach. It involves a fixed point $F=(h,k) \in \mathbb{R}^2$ called the \textit{focus}, a line $L \in \mathbb{R}^2$  not containing $F$ called the \textit{directrix}, and a nonnegative number $e$ called the \textit{eccentricity} (not to be confused with the Euler number $e$). A conic section is defined as the locus of all points for which the ratio of their distance to $F$ to their distance to $L$, is equal to $e$. We obtain an ellipse when $e \in [0,1)$, a parabola when $e=1$, and a hyperbola when $e>1$. A circle is an  ellipse with $e=0$. See \cite{Akopyan_Zaslavsky_2007} and \cite{Pierre_1988} for a comprehensive review of the basic concepts of~conics. See Figure \ref{F:plot100} for a graphical illustration.

Denote by $M$ the line containing the axis of symmetry in the case of a hyperbola or parabola, and containing the major axis in the case of an ellipse. Line $M$ contains the focus $F$ in all three types of conic sections. The  \textit{rotation angle} of the conic is defined as the counterclockwise angle, $\varphi$, from the positive X-axis to  the line~$M$. Given a conic point, $\boldsymbol{w}_i=(u_i, v_i)$, denote  by $S_i$ the line segment connecting this point to the focus. Then the point $\boldsymbol{w}_i$ satisfies the equation
\begin{equation}
r_i = \frac{l}{1+e \cos t_i}, \label{r_i}
\end{equation}
where $r_i$ is the length of the segment $S_i$, $t_i$ is the counterclockwise angle from the line~$M$ to the segment $S_i$, and
$l$ is a positive constant known as the \textit{semi-latus rectum}. Figure \ref{F:plot100} graphically illustrates these concepts. Due to the radial symmetry of a circle, the line $M$ is not uniquely defined, and can be arbitrarily taken to be parallel to the positive X-axis. 

The angles  $t_1,\ldots,t_n$ belong to the interval $[-R(e),R(e)]$, where
\begin{align}
R(e) =
\begin{cases}
\pi \quad &\text{for ellipses and parabolas},\\
\arccos(-e^{-1}) \quad &\text{for hyperbolas}.\\
\end{cases} \label{e_support}
\end{align}
Consequently, for ellipses and parabolas,  angle $t_i$ is unrestricted and belongs to the interval $(-\pi,\pi)$. For hyperbolas, angle $t_i$ belongs to the interval $\bigl(-\arccos(-e^{-1}),$ $\arccos(-e^{-1})\bigr)$.
The vector of hyperparameters uniquely determining the conic section is  $\boldsymbol{\theta}=(h,k,\varphi, l, e) \in \Theta$. Distance $r_i$ in equation (\ref{r_i}) is a function of  parameter $\boldsymbol{\theta}$ and angle $t_i$. Transforming from polar to Cartesian coordinates, and explicitly specifying the dependence on the model parameters,
we obtain the following equations for any conic point, $\boldsymbol{w}_i=(u_i, v_i)$:
\begin{align}
u_i &= u(t_i, \boldsymbol{\theta}) = h + r_i(t_i, \boldsymbol{\theta}) \cdot\cos(t_i+\varphi), \label{xy_i}\\
v_i &= v(t_i, \boldsymbol{\theta}) = k + r_i(t_i, \boldsymbol{\theta}) \cdot\sin(t_i+\varphi). \notag
\end{align}

\subsection{Quadratic equation definition using Cartesian coordinates}\label{S:quad equation}

 An equivalent and more popular formula for a conic point  is
\begin{equation}
Au_i^2 + 2Bu_i v_i + Cv_i^2 + 2Du_i + 2Ev_i + F = 0, \label{conic}
\end{equation}
where $A$, $B$ and $C$ are not simultaneously zero. Without loss of generality, and for identifiability of these parameters, we may assume that $A^2+B^2+C^2=1$;  other restrictions are also available in the literature, e.g., see \citep{Zhang_1997}. We denote the vector $\boldsymbol{\vartheta}=(A,B,C,D,E,F)$ as representing the parameters of the conic section. Notice that due to the   restriction,  $\boldsymbol{\vartheta}$ lies in a 5-dimensional space. It can be shown that there is a one-to-one map between the parameter $\boldsymbol{\theta}$ of the focus-directrix formulation  and  parameter $\boldsymbol{\vartheta}$ of the  algebraic formulation.
Defining the constant $I=A+C$, and the matrix determinants
\begin{equation*}
\Delta=
\left|
\begin{matrix}
A &B &D\\
B  &C &E\\
D &E &F\\
\end{matrix}
\right| \quad\text{and}\quad
J=
\left|
\begin{matrix}
A &B\\
B  &C\\
\end{matrix}
\right|,
\end{equation*}
the additional assumptions that $\Delta \neq 0$, $J > 0$, and $\Delta/I < 0$ guarantee that  equation~(\ref{conic})  represents an ellipse rather than a hyperbola or parabola. Let $\Upsilon$ be the parameter space corresponding to these additional constraints on  vector $\boldsymbol{\vartheta}$. It follows that there exists a one-to-one mapping between elements of the  parameter space $\Upsilon$ and the parameter space $\Theta$ defined in Section \ref{S:FD} \citep[e.g., see][]{Silverman_2012}. Thus, for statistical estimation we can work with either parametrization. Finally, we present the formulation of conics in the most popular standard form by centering and rotation.

\subsection{Conics in standard form}\label{S:standard}

We define the \textit{center} of a conic as follows. For an ellipse, it is the midpoint of the segment joining the two foci. For a parabola, the center is the point at which the axis of symmetry intersects the parabola. For a hyperbola, it is midway between the foci of the two branches. A conic in \textit{standard form} has its center at the origin and rotational angle $\varphi=0$.

\paragraph{Identifiability of angle $\varphi$ in standard form} We assume that in standard form, parabolas are left-opening,   ellipses have the horizontal axis as their major axes, and we consider only the left branch of hyperbolas. Right-opening hyperbolas are regarded as rotated versions of left-opening hyperbolas with angle $\varphi=\pi$. These assumptions ensures that rotational parameter $\varphi$ is identifiable when estimating this parameter based on noisy data. The equations for  standard conics are then given for $a, b>0$:
\begin{description}
\item[\textit{Ellipses:}] A standard ellipse  has the equation $u^2/a^2 + v^2/b^2 = 1$.

\item[\textit{Hyperbolas:}] A (left-opening) standard hyperbola has the equation $u^2/a^2 - v^2/b^2 = 1$.

\item[\textit{Parabola:}] A left-opening standard parabola has the equation $v^2 =- 4 a u$.
\end{description}
\smallskip

\noindent Notice that the focus $F=(h,k)$ is different from the conic center and  has a  different expression for each kind of  conic in standard form. Whereas the center coincides with the origin, the expression for $F$ depends on the constant $a$ in the case of parabolas, and on the constants $a$ and $b$ in the case of  ellipses and hyperbolas.

\paragraph{Converting a conic equation to standard form}

Let $\boldsymbol{w}=(u, v)$ be a  point lying on an arbitrary (non-standard) conic whose  center is $(c_1,c_2)$. We make the  transformation:

\begin{align}
\left(
\begin{matrix}
u_{rs}\\
v_{rs}\\
\end{matrix}
\right)
=
\left(
\begin{matrix}
\cos\varphi &\sin\varphi\\
-\sin\varphi &\cos\varphi\\
\end{matrix}
\right)
\left(
\begin{matrix}
u-c_1\\
v-c_2\\
\end{matrix}
\right)
\label{rs}
\end{align}
Then the conic equation expressed in terms of $(u_{rs},v_{rs})$ is in standard form.

\subsection{Estimation of conic parameters with known type}\label{S:lit review}
There is  vast literature on numerical methods for the estimation of parameters of a given conic section when it is known that the (noisy) data arise from one of the three basic types of conic sections. For instance, \cite{Zhang_1997} presents a very comprehensive review of several commonly used techniques, including  linear least-squares (pseudo-inverse and eigen analysis), orthogonal least-squares, gradient-weighted least-squares, and  bias-corrected renormalization.
\cite{Prasad_etal_2013} proposed a method specifically designed for  fitting  ellipses using the geometric distances of the ellipse from the data points.
Given a specific type of conics, \cite{Ahn_etal_2001} provide various estimation methods for different types of conics sections using  least-squares with respect to various predefined measures.

However, none of these above mentioned methodologies are able to provide probabilistic inference about the underlying hidden type of a conic section. Moreover, as the majority of  these earlier methodologies are based on some form of optimization (that minimizes a chosen distance between the observed points and the known type of conic section), uncertainty estimates for the conic parameters are not readily available. Thus, full statistical inference including standard errors of estimates is often missing in some of these earlier works. Bootstrap methods could perhaps be used to obtain standard errors of these earlier estimates, but that would require establishing asymptotic theory of the underlying data measured with errors when the support of the true data is a given conic section.

We instead adopt a model-based, fully Bayesian hierarchical approach using the focus-directrix formulation, which facilitates the complete probabilistic inference of not only the parameters for a given type of conic section, but also provides the posterior probability of classification for the conic section type. The proposed estimation technique relies on a Bayesian computational approach and the focus-directrix representation to model the data and obtain the entire posterior distribution of the conics parameters. It has several practical and methodological advantages that make it an important addition to the aggregation of existing inferential techniques for conics.

The advantages the proposed method include the ability to: \textit{(i)} provide a coherent, integrated framework for flexibly fitting different conic sections (hyperbola, parabola, ellipse, and circle), \textit{(ii)}  make accurate inferences on the unknown type of conic section underlying the data, \textit{(iii)} compute uncertainty estimates for all the model parameters, and \textit{(iv)} achieve high fidelity to true conics even in challenging inferential situations, such as noisy data arising from partial conics in arbitrarily rotated and non-standard form. As the simulation studies in  Section \ref{S:simulation} demonstrate, the technique is impressively accurate even in noisy, partial datasets like Figure \ref{F:plot0}, where a visual inspection is unable to correctly call the hidden type of conic section underlying the data.

The rest of the paper is organized as follows. Section \ref{S:model} specifies the Bayesian model and Section \ref{S:inference} outlines the inference procedure. Section \ref{S:simulation} demonstrates the accuracy of the proposed technique using simulation studies and makes comparisons with some existing techniques.

\section{A Bayesian Inferential Framework for Conics}\label{Bayesian}

\subsection{The hierarchical model}\label{S:model}

Bayesian models typically consist of a sampling (conditional) distribution of the data given the possibly vector-valued model parameter. This contributes towards the calculation of the likelihood function of the parameter. Additionally, the model requires specification of the marginal probability distribution for the parameter, which is commonly known as the prior distribution. Then, using the Bayes' Theorem, we obtain the conditional distribution of the parameter given the observed data, commonly known as the posterior distribution. For a majority of the Bayesian models, it is  impossible to obtain the analytical form of the posterior distribution.  Markov Chain Monte Carlo (MCMC) methods are then utilized to draw approximate samples from the posterior distribution \citep[e.g.,][]{Gelman_etal_2014}.

\paragraph{Likelihood function} The noisy observations are modeled as  arising from a set of latent conic points contaminated by independent Gaussian errors. In other words, referring back to the focus-directrix approach of Section \ref{S:FD}, the observations can be represented as:
\begin{align}
x_i &= u(t_i, \boldsymbol{\theta}) + \epsilon_{i1}, \notag\\
y_i &= v(t_i, \boldsymbol{\theta}) + \epsilon_{i2}, \quad\text{where} \notag\\
\epsilon_{i1}, \epsilon_{i2} &\stackrel{iid}\sim N(0,\sigma^2) \label{likelihood1}
\end{align}
for $i=1,\ldots,n$ and with an unknown  $\sigma>0$ denoting the measurement error variance. It is possible to relax the assumptions on the error by extending it to a bivariate distribution that allows for correlated errors, but in order to keep the exposition simple, we work with the i.i.d.\ case. However, in real applications, one may wish to check the distributional assumption of the errors by an appropriate suite of goodness-of-fit tests \citep{HuberCarol_eta_2002}. The Cartesian coordinates of the latent conic point is given in equation~(\ref{xy_i}). Let $\boldsymbol{t}=(t_1,\ldots,t_n)$ be the set of (unobserved) angular parameters and parameter vector $\boldsymbol{\eta}=(\boldsymbol{\theta},\sigma^2)$ be the full set of hyperparameters. Conditional on the latent angular variables, the model implies the following (conditional) joint density function for the data $\mathcal{W}$ given the angular variables:
\begin{equation}
[\mathcal{W} \mid \boldsymbol{t}, \boldsymbol{\eta}] = \prod_{i=1}^n N\bigl(x_i \mid u(t_i, \boldsymbol{\theta}), \sigma^2\bigr) \cdot \prod_{i=1}^n N\bigl(y_i \mid v(t_i, \boldsymbol{\theta}), \sigma^2\bigr)\label{likelihood}
\end{equation}
where the symbol $N(x \mid \mu, \sigma^2)$ represents the probability density function of a normal distribution with mean $\mu$ and variance $\sigma^2$ evaluated at the point $x$. Later, we provide details on how we model the latent angular parameters $t_i$. Next, we specify the prior probability distributions of the parameters of the conic sections.

\paragraph{Eccentricity}

This parameter is key because it identifies the type of conic section (ellipse, parabola, or hyperbola) underlying the data. In the absence of any prior information about the relative abundances of the three conics, it is reasonable to assume that the three conics types are equally likely. That is, $P(0\le e<1)=$ $P(e=1)=$ $P(e > 1)=$ $1/3$. Clearly, we can not use any probability distribution on $e$ that is dominated by a Lebesgue measure as it will not allow for a positive probability for circles and parabolas. Thus, we need to use a slightly non-standard prior for $e$  that  allows positive probabilities for the events $\{e=0\}$ and $\{e=1\}$, facilitating posterior inferences about this important parameter.

Let $F_0(\cdot)$ be a continuous CDF with support $(0,\infty)$ and median greater than 1. That is, $F_0(0)=0$, $F_0(1)<1/2$, and $F_0(\infty)=1$. 
We specify the prior for eccentricity parameter $e$ as follows:
\begin{align}
e
\begin{cases}
=0 \quad&\text{w.p. $\frac{1-2F_0(1)}{3(1-F_0(1))}$}, \\
=1 \quad&\text{w.p. $\frac{1}{3}$}, \\
\sim F_0 \quad&\text{w.p. $\frac{1}{3(1-F_0(1))}$},\\
\end{cases}\label{e2}
\end{align}
with the last branch corresponding to the condition $e \in (0,\infty)\backslash\{1\}$, i.e., non-circular ellipses as well as hyperbolas.
Prior  (\ref{e2}) implies that $P(0 \le e < 1)=$ $P( e = 1)=$ $P( e > 1)=$  $1/3$, so that ellipses, parabolas, and hyperbolas are apriori equally likely, yielding a mixture prior on the different types of conics.

We allow the user to choose suitable prior distributions within the above set up, and later through simulation studies, demonstrate that the posterior inferences are relatively insensitive to such prior choices.
Specifically, if  distribution $F_0$ is such that $F_0(1)=9/19 < 1/2$ (e.g., exponential distribution with mean 1.557),
we  obtain the following prior for the eccentricity parameter:
\begin{align}
e
\begin{cases}
=0 \quad&\text{w.p. $1/30$}, \\
=1 \quad&\text{w.p. $1/3$}, \\
\sim F_0 \quad&\text{w.p. $19/30$}.\\
\end{cases}\label{e}
\end{align}

Posterior inferences are made about the probability that the  conic section underlying the noisy data is  a particular type, e.g., parabola. The posterior probabilities in Table \ref{T:simulation011} were computed in this manner.
If interest focuses on  the Bayes factor for a conic type,  this can be easily evaluated from the posterior probabilities. For example, consider the data generated by Parabola \#38 in Table  \ref{T:simulation011}. The estimated posterior probability of the event $[e=1]$ is $0.995$, corresponding to a posterior odds of $0.995/(1-0.995)=199$. The prior odds is $\frac{1/3}{2/3}=0.5$. The Bayes factor in favor of the hypothesis that the  underlying conic is a parabola, versus the alternative hypothesis that it is not a parabola, is therefore the ratio of the posterior odds to the prior odds, i.e., $199/0.5=398$. This is overwhelming evidence in favor of parabolas. For Hyperbola \#9 and Ellipse \#90, the Bayes factors in favor of the true conics are computationally infinite.

\paragraph{Angular parameters} Let $t_i^*$ be a  version of  angle $t_i$  standardized to the unit interval. Specifically, let
\begin{equation}
t_i^* =
\begin{cases}
(t_i + \pi)/\bigl(2\pi\bigr) \quad&\text{if $e\le 1$}\\
\bigl(t_i + \arccos(-e^{-1})\bigr)/\bigl(2\arccos(-e^{-1})\bigr) \quad&\text{otherwise.}
\end{cases}\label{std_ti}
\end{equation}
Clearly, prior to being given the noisy data, we do not have any strong
information about the distributional shape of the angular variables, and so it is reasonable to model these latent
variables using a flexible class of probability distributions that would not impact the posterior inference,  and would additionally allow  arbitrary shapes for the density functions (e.g., symmetric, skewed, bimodal).

We adopt a mixture of (a known sequence of) Beta densities which has been shown to approximate any continuous density supported on the unit interval $[0, 1]$ \citep{Vitale_1975, Babu_etal_2002, Ghosal_2001}. Given any continuous density $f(t)$ supported on $[0, 1]$, it can be shown that $\sum_{s=1}^{m+1} p_s \text{beta}(s,m+2-s)$ converges uniformly to $f(t)$ as $m\rightarrow\infty$ if we choose $\tilde{p}_s=f((s-1)/(m+2))$ for $s=1,\ldots,m+1$ and define $p_s=\tilde{p_s}/\sum_s\tilde{p_s}$. Motivated by this uniform convergence result, for a prespecified positive integer $m \ge 4$ that is typically chosen to be large, we model the standardized angles $t_1^*,\ldots,t_n^*$ as being apriori exchangeable and having a mixture of beta distributions:
\begin{align}
t_i^* &\stackrel{iid}\sim \sum_{s=1}^{m+1} p_s \, \text{beta}(s,m+2-s), \quad\text{where probability vector}\notag\\
\boldsymbol{p}=(p_1,\ldots,p_{m+1}) &\sim \mathcal{D}_{m+1}(\alpha,\cdots,\alpha). \label{p}
\end{align}
Here, $\mathcal{D}_{m+1}(\alpha,\cdots,\alpha)$ denotes a Dirichlet distribution on $m+1$ categories with the  concentration parameters all equal to $\alpha>0$. Prespecifying $\alpha=1$ gives satisfactory results in most applications.

In addition to  simplicity of implementation, this  framework has important advantages. Being a linear combination of Bernstein basis polynomials of degree $m$, the prior density is able to closely approximate any true continuous density of the angular parameters, provided $m$ is large enough \citep[e.g.][]{Natanson_1964}. 
In fairly general situations, applying an asymptotic result of \cite{Babu_etal_2002}, \cite{Turnbull_Ghosh_2014}  recommend setting the degree $m=\lceil\log n / n\rceil$.

\paragraph{Hyperparameters}

The eccentricity $e$, semi-latus rectum $l$, and  center coordinates $h$ and $k$ are all assigned independent Gaussian prior distributions, restricted to the positive real line for the  parameters that are positive. The rotational angle $\varphi$ is assigned a uniform prior on the interval $(-\pi,\pi]$. Variance  $\sigma^2$ is given an inverse gamma prior.

\smallskip

\subsection{Posterior inferences}\label{S:inference}

Initial estimates for the model parameters are computed as described in Section \ref{S:initialization}. With these estimates as the starting values,
the model parameters are  iteratively updated using the  MCMC steps outlined in Section \ref{S:MCMC}. Subsequently, the post--burn-in MCMC sample is used to make posterior inferences.

\subsubsection{Initialization procedure for MCMC algorithm}\label{S:initialization}

We first present a basic result:

\begin{lemma}\label{S:lemma}
Let $\boldsymbol{w}=(u, v)$ be a  point on the conic with center $\boldsymbol{c}=(c_1,c_2)$. Transform $\boldsymbol{w}=(u, v)$ to obtain  point $\boldsymbol{w}^*=(u^*, v^*)$ as follows:
\begin{equation}
\left(
\begin{matrix}
u^*\\
v^*\\
\end{matrix}
\right)
=
\left(
\begin{matrix}
\cos\varphi &\sin\varphi\\
-\sin\varphi &\cos\varphi\\
\end{matrix}
\right)
\left(
\begin{matrix}
u\\
v\\
\end{matrix}
\right) \label{12A} 
\end{equation}
This  transformation  rotates the conic point  clockwise around the origin by an angle of $\varphi$. Similarly,
 center $\boldsymbol{c}=(c_1,c_2)$ is   premultiplied by the  square matrix appearing in equation (\ref{12A}) to obtain the rotated center, $\boldsymbol{c}^*=(c_1^*,c_2^*)$.

\begin{enumerate}
\item\label{L:elps_hypl} Suppose the conic section is either an ellipse ($\delta=1$) or hyperbola ($\delta=-1$).  Then
    \begin{equation}
    {v^*}^2 - \beta_0 - \beta_1u^*- \beta_2{u^*}^2 - \beta_3v^*=0, \label{elps_hypl}
     \end{equation}
     where $\beta_3=2c_2^*$, $\beta_2=-\delta b^2/a^2$, $\beta_1=2\delta c_1^*b^2/a^2$, and $\beta_0=-{c_2^*}^2-\delta {c_1^*}^2b^2/a^2+\delta b^2$.

    \item\label{L:prbl} Suppose the conic section is a parabola. Then
    \begin{equation}{v^*}^2 - \beta_0 - \beta_1u^*- \beta_2v^*=0, \label{prbl}
     \end{equation}
     where $\beta_1=-4a$, $\beta_2=2c_2^*$, and $\beta_0=-{c_2^*}^2+4ac_1^*$.

\end{enumerate}
\end{lemma}

The proof of the Lemma uses equation (\ref{rs}), which converts the conic equation to standard form, and involves making use of  standard well-known results \citep[e.g.\ ][pp.\ 44--45]{Fanchi_2006} and  some tedious algebra that we have omitted for brevity.  

\smallskip

  The starting parameters are computed as follows:

\begin{enumerate}
\item Using the quadratic equation representation of conics (see Section \ref{S:quad equation}), we apply the least-squares fitting approach of \citep[Sec.\ 4.2]{Zhang_1997} to compute a preliminary estimate,  $\hat{\boldsymbol{\vartheta}}=$ $(\hat{A},\hat{B},\hat{C},\hat{D},\hat{E},\hat{F})'$, of the conic parameters.

    \item It can be shown that the true rotational angle, $\varphi$, satisfies
    $
    \tan (2\varphi) = \frac{2B}{A-C}
    $. Let $\hat{\varphi}_1=$ $\frac{1}{2}\arctan(\frac{2\hat{B}}{\hat{A}-\hat{C}})$,  so that $\hat{\varphi}_1 \in (-\pi/2,\pi/2)$. Possible  estimates for the true unknown $\varphi$, which may belong to $(-\pi,\pi]$, are then contained in the set
     \[
     \{\lfloor\hat{\varphi}_j\rfloor: \hat{\varphi}_j=\hat{\varphi}_1+(j-1)\,\pi, \, j=1,2,3,4\}.
      \]
      with the symbol $\lfloor\cdot\rfloor$ denoting a wrapped angle belonging to the interval $(-\pi,\pi]$, e.g., $\lfloor9\pi/4\rfloor=\pi/4$.

\item For each of the four candidate estimates, $\lfloor\hat{\varphi}_j\rfloor$:
\begin{enumerate}

\item Use  angle $\lfloor\hat{\varphi}_j\rfloor$ to rotate the data point $(x_i,y_i)$ clockwise around the origin and thereby obtain the point $(x^*_{i},y^*_{i})$. Similar to  equation~(\ref{12A}), this is done by premultiplying vector $(x_i,y_i)$ by the square matrix determined by the angle $\lfloor\hat{\varphi}_j\rfloor$.

    \bigskip

    \item\label{init_elps} \textbf{Assuming that the conic is an  ellipse}:
    \begin{enumerate}

    \item\label{theta_init_elps} Based on expression (\ref{elps_hypl}), we find an estimate $\hat{\boldsymbol{\beta}}=(\hat{\beta}_0,\hat{\beta}_1,\hat{\beta}_2,\hat{\beta}_3)$ that minimizes the quantity $\sum_{i=1}^n\bigl({y_i^*}^2 - \beta_0 - \beta_1x_i^*- \beta_2{x_i^*}^2 - \beta_3y_i^*\bigr)^2$. This is equivalent to computing the least squares estimate of $\boldsymbol{\beta}$ in the model ${y_i^*}^2 = \beta_0 + \beta_1x_i^* + \beta_2{x_i^*}^2 + \beta_3y_i^* + \epsilon_i$, where the independent errors $\epsilon_i$ have zero means and equal variances. Using these estimates, we  apply Part (\ref{L:elps_hypl}) of Lemma \ref{S:lemma} to estimate the ellipse parameters $(\hat{a},\hat{b},\hat{h},\hat{k})$, which is in one-to-one correspondence with the estimated parameters  of the focus-directrix approach, denoted by $\hat{\boldsymbol{\theta}}_j^{(-)}$.

        \item\label{ti_init_elps} Applying expression (\ref{xy_i}) of the present paper, we  express the \textit{nearest}  conic point to the $i^{th}$ datum  $\boldsymbol{w}_i=(x_i,y_i)$ as $\boldsymbol{w}(\hat{t}_i)=\bigl(u(\hat{t}_i), v(\hat{t}_i)\bigr)$ for some optimal angle $\hat{t}_i$. In other words,
    \[
\hat{t}_i = \text{argmin}_{l_j \le t< u_j}  \bigl\| \boldsymbol{w}_i - \boldsymbol{w}(t) \bigr\|
\]
where expression (\ref{e_support}) of the paper gives the support $(l_j,u_j) =
(-\pi,\pi)$ for an ellipse.
    Consequently, the initial estimate $\hat{t}_i$ can be found by a univariate optimizer over the interval $(-\pi,\pi)$. This procedure is repeated for the $n$ data points to get an initial estimate $\hat{\boldsymbol{t}}$ for the vector of $n$ angular parameters.

\item Expression (\ref{likelihood}) is maximized over  $\sigma>0$ to obtain the likelihood function for the best-fitting ellipse corresponding to rotational angle $\lfloor\hat{\varphi}_j\rfloor$. Let the likelihood function for the best-fitting ellipse be denoted by $l_{j}^{(-)}$.
    \end{enumerate}

\bigskip

 \item\label{init_hybl} \textbf{Assuming that the conic is a hyperbola}, we perform a similar initialization procedure to Part \ref{init_elps}. The only difference is with Part \ref{ti_init_elps} in which, applying expression (\ref{xy_i}) of the main paper, the support of the univariate optimizer for the angles of a hyperbola is $(l_j,u_j) =
\bigl(-\arccos(-1/\hat{e}_j^{(+)}),$ $\arccos(-1/\hat{e}_j^{(+)})\bigr)$ where $\hat{e}_j^{(+)}$ is the eccentricity parameter estimated as part of parameter vector $\hat{\boldsymbol{\theta}}_j^{(+)}$ for the focus-directrix approach. We thereby obtain the likelihood function, $l_{j}^{(+)}$, for the best-fitting hyperbola corresponding to the rotational angle $\lfloor\hat{\varphi}_j\rfloor$.

\bigskip

\item\label{init_prbl} \textbf{Assuming that the conic is a parabola}, we perform a similar initialization procedure to Part \ref{init_elps}. The only difference is in Part \ref{theta_init_elps}, in which we instead apply expression (\ref{prbl}) in Part (\ref{L:prbl}) of Lemma \ref{S:lemma} to obtain the estimated parameter $\hat{\boldsymbol{\theta}}_j^{(0)}$ for the parabola. In this manner, we obtain the likelihood function, $l_{j}^{(0)}$, for the best-fitting parabola corresponding to the rotational angle $\lfloor\hat{\varphi}_j\rfloor$.

\bigskip

\item Compute the overall maximized likelihood over the conic types for the rotational angle $\lfloor\hat{\varphi}_j\rfloor$:
\[
l_j^{\max} = \max\bigl\{l_{j}^{(-)}, l_{j}^{(0)}, l_{j}^{(+)}\bigr\}
\]
\end{enumerate}

\item Set the estimated rotational angle to the candidate angle with the greatest overall maximized likelihood. That is, $\hat{\varphi} = \lfloor\hat{\varphi}_{j^*}\rfloor$ where $j^* =\underset{j=1,2,3,4}{\text{argmax}}\, l_j^{\max}$.

\item For the estimated rotational angle $\hat{\varphi}$, find the most likely conic type and its associated parameters as described in Steps \ref{init_elps}, \ref{init_hybl}, and \ref{init_prbl}.
\end{enumerate}

\subsubsection{MCMC updates}\label{S:MCMC}

With the estimates from the previous section as the initial values,
the model parameters are  iteratively updated using the following MCMC algorithm.

\paragraph{Eccentricity parameter} For conditionally updating the parameter $e$,
it can be shown that the upper bound for $e$ equals $-1/\cos(\max_{i} |t_i|)$ if $\max_{i} |t_i|$ exceeds $\pi/2$, and equals $\infty$ otherwise \citep[e.g.,][]{Silverman_2012}. Ideally, we would like to compute the posterior probabilities of the three branches of the prior~(\ref{e}). However, the third branch involves  a marginal likelihood calculation for a non-conjugate part of the model. We therefore apply the Laplace approximation to obtain an approximate, conjugate normal model with restricted support, for which the marginal likelihood is available in computationally closed form. This approximate model is used to propose a new parameter value that is either accepted or rejected with a Metropolis-Hastings (MH)  probability to obtain draws from the true conditional posterior of the parameter $e$.

\paragraph{Inferred type of conic section} We  define a categorical variable identifying the type of conic section from the eccentricity parameter:
\begin{align}
\mathcal{C}=
\begin{cases}
\text{circle} \quad &e = 0\\
\text{non-circular ellipse} \quad &0 < e < 1\\
\text{parabola} \quad &e =1\\
\text{hyperbola} \quad &e > 1\\
\end{cases}\label{detectingConic}
\end{align}
The detected type of conic section is the mode of the post--burn-in MCMC samples of this categorical variable. Empirical average estimates of the posterior probabilities of the four categories, along with their uncertainty estimates, are also readily available from the MCMC sample. The results  displayed in Table \ref{T:simulation011} were computed in this manner.

\paragraph{Remaining model parameters} The hyperparameters $(h,k)$, $l$, and $\sigma^2$ can be updated by Gibbs sampling because of the model's conditional conjugacy in these parameters.
Parameters $\varphi$ and $t_1,\ldots,t_n$ are conditionally updated by random walk MH moves. For updating these parameters, the normal proposal distribution's variance is set equal to a constant times the inverse Fisher information matrix, with the constant chosen to  achieve an empirical  acceptance rate of    25\% to 40\% for the MH proposals. For obtaining fast-mixing MCMC chains, this is the recommended range of acceptance rates  in overdispersed random walk MH algorithms  \citep{Roberts_etal_1997}. Due to relation (\ref{e_support}),
the normal proposals for the angle parameters $t_1,\ldots,t_n$ are restricted to the interval $(-\pi,\pi)$ for ellipses and parabolas, and to the interval $\bigl(-\arccos(-e^{-1}),\arccos(-e^{-1})\bigr)$ for~hyperbolas. Finally, the relative weights of the beta prior distributions in expression (\ref{p}) are updated using the current values of the angles $t_1,\ldots,t_n$.

\subsubsection{Posterior Estimation and credible intervals for the true  conic}\label{S:HPD}

Recall that, in the focus-directrix approach, hyperparameter $\boldsymbol{\theta}=(h,k,\varphi,$$ l, e)$ uniquely determines the conic from which the noisy data have been generated.
Using the post--burn-in MCMC sample, the empirical average  of the sampled values of vector $\boldsymbol{\theta}$ provides  a point estimate for the  true  conic section. The  estimated curves in Figure \ref{F:plot0}, displayed by the blue bold lines, were computed in this manner.

Uncertainty estimates for the true   conic  are also available. Suppose we are interested in a $100(1-\gamma)$\% posterior credible region (CR) for the true conic, for some $0<\gamma<1$. Using the MCMC sample, we compute   $100(1-\frac{\gamma}{5})$\% marginal posterior credible intervals (CIs) for each   component of hyperparameter $\boldsymbol{\theta}=(h,k,\varphi, l, e)$. Then the Cartesian product of those five CIs is a conservative $100(1-\gamma)$\% posterior CR for the true  conic. In Figure \ref{F:plot0}, the gray lines represent conic samples generated from conservative 95\% CRs of this type. We find from Figure \ref{F:plot0} that these curves are good fits for the data, even in the case of the ellipse, where the sampled conics  provide appropriate coverage for the noisy data points.

Since hyperparameter $\boldsymbol{\theta}$ is not high-dimensional, we do not expect the Bonferroni-type adjustments for the CIs to be too conservative in general. However, this is certainly possible in some studies, especially when the components of $\boldsymbol{\theta}$ are highly correlated in the posterior. In these situations, to overcome the issue of marginalizing over the angular parameters $t_1,\ldots,t_n$ and variance $\sigma^2$, we recommend the following computational strategy to avoid an overly conservative CR. The MCMC sample is post-processed to estimate the posterior mean and variance matrix of $\boldsymbol{\theta}$. A sample of $\boldsymbol{\theta}$ values is generated from the 5-variate normal distribution having this mean  and variance, which represents a large-sample approximation to the marginal posterior density of $\boldsymbol{\theta}$. Evaluating the approximately normal densities for these sampled $\boldsymbol{\theta}$, and retaining only those belonging to the top 95th percentile of   densities, gives an approximate 95\% highest posterior density CR for the true conic.

\bigskip

\section{Numerical illustrations using simulated data scenarios}\label{S:simulation}

The artificial datasets in Section \ref{S:ellipse simulation} were generated  using the simulation strategy of  \cite{Zhang_1997}. We investigate the accuracy with which we are able to infer the ellipse parameters from noisy data, including estimation of the rotational angle $\varphi$. Artificial data were generated from  partial as well as complete true ellipses. In Section \ref{S:ellipse simulation}, we  assume that the type of conic section, e.g. ellipse, is known, and focus on estimating the ellipse parameters. We  relax this assumption in the next set of simulations in Section \ref{S:simulation_1}, generating data from partial ellipses, parabolas, and hyperbolas having randomly generated rotational angles and other model parameters.

\subsection{Simulation Study 1}\label{S:ellipse simulation}

We  investigated the ability of the Bayesian  procedure to detect standard-form ellipses   from noisy data.
Similar to the  strategy of Section \ref{S:simulation_1} and Section 10.1 of \cite{Zhang_1997}, we assumed that the true conic is an ellipse with the following parameters: the major axis equals $2a=100$ units, the
minor axis equals $2b=50$ units, the center is at $(c_1, c_2)=(250,250)$,
and the rotation angle is 0 degrees. 

We generated $n=200$ data points with the standardized angular parameters, defined in equation (\ref{std_ti}), generated as  $t_i^* \stackrel{iid}\sim \text{beta}(3,3)$ for $i=1,\ldots,n$. From these values, the angular parameters $t_i=2\pi \,t_i^*-\pi$ were computed. A rotational angle of $\varphi=\pi$ was assumed. A bivariate  error with independent Gaussian  components having mean 0 and standard  deviation $\sigma=2$ was added to each  ellipse point to obtain the corresponding data point.

\begin{figure}[h!]
\begin{center}
\vspace{-10pt}
\includegraphics[scale=0.5]{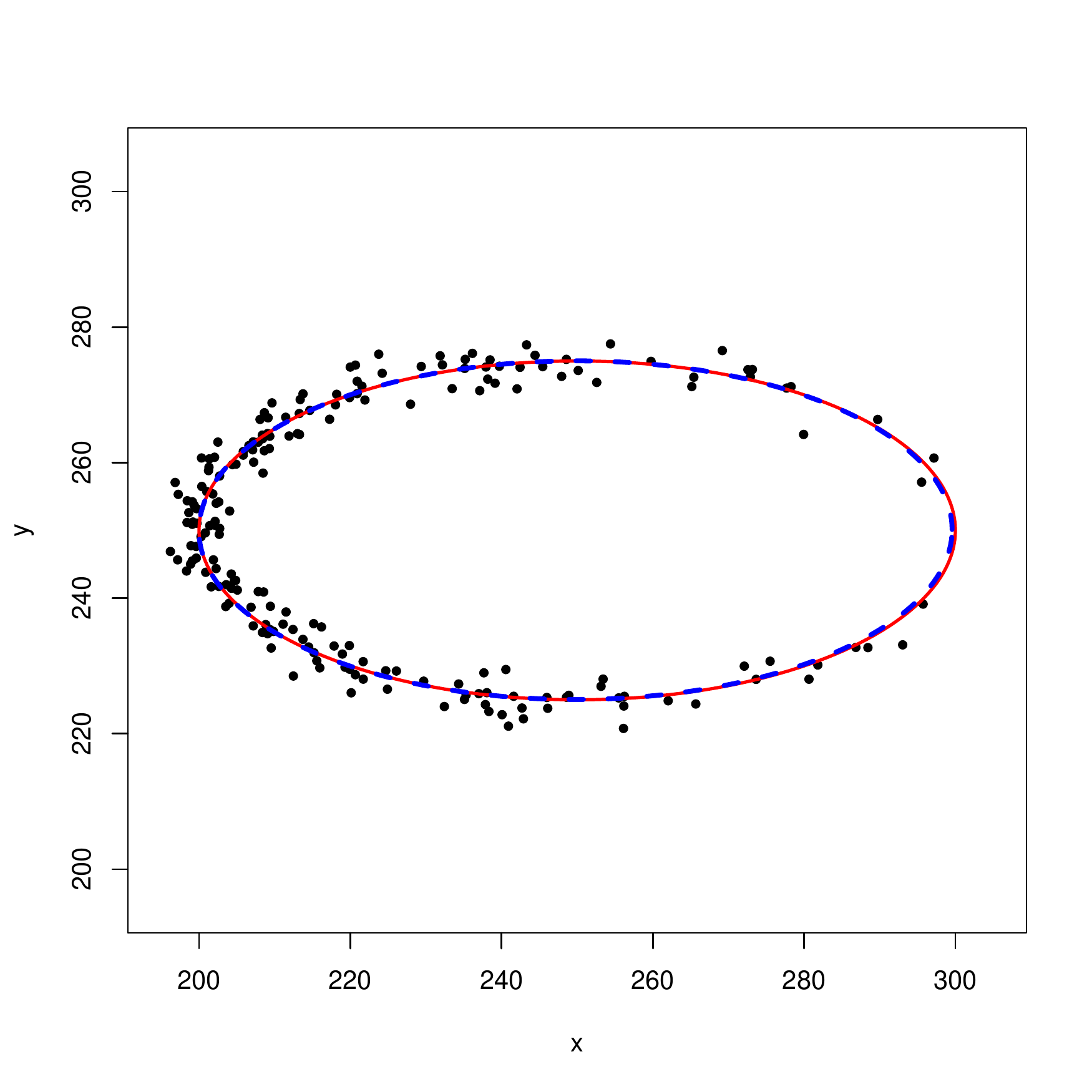}
\includegraphics[scale=0.5]{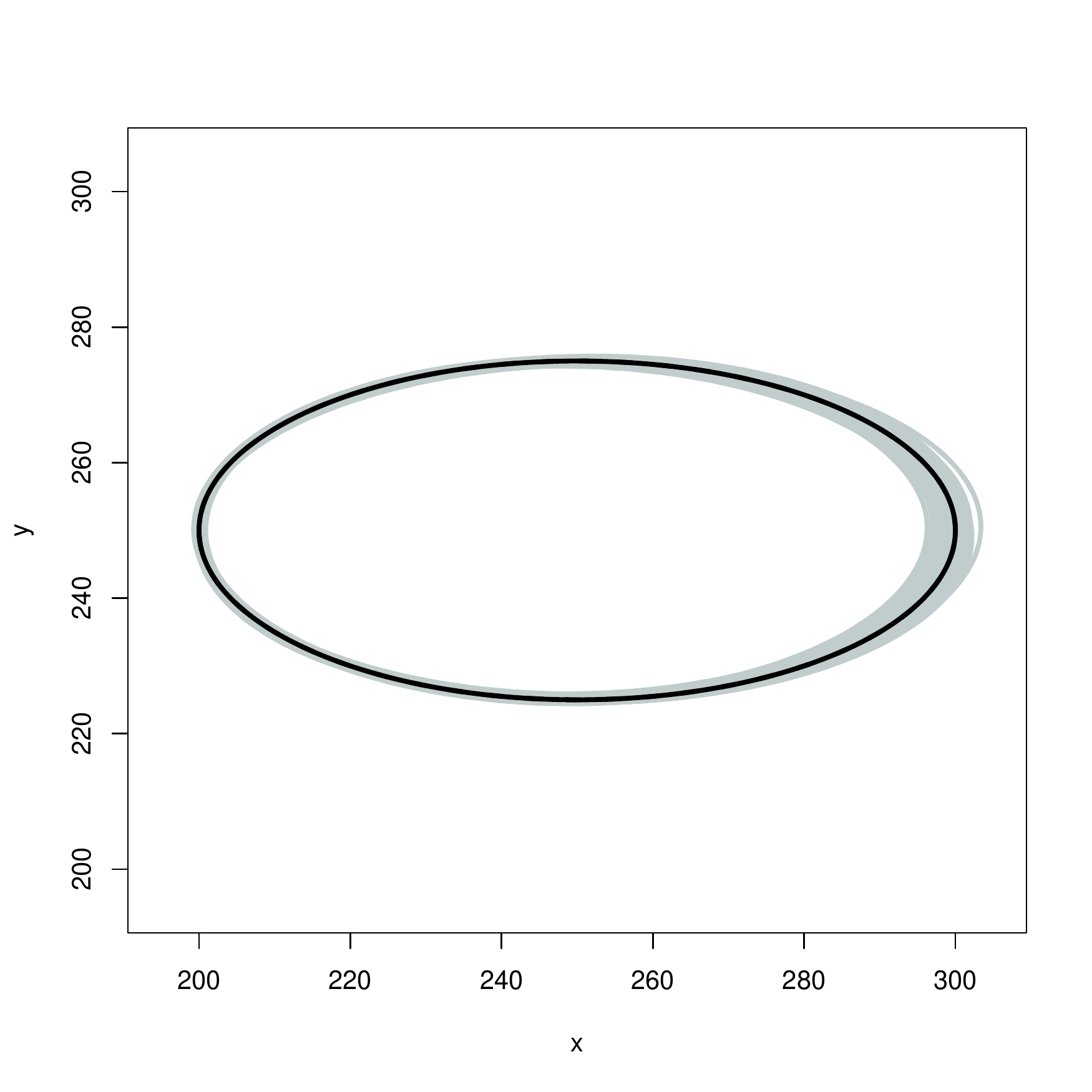}
\caption{For a typical dataset in Simulation 1, the top panel plots the data points using black dots. The red solid line represents the true ellipse. The blue dashed line is the estimated ellipse for the Bayesian approach with inferred parameters averaged over the 100 datasets. The lower panel displays the true ellipse (black) and the 100 estimated ellipses (gray) from the artificial datasets for the Bayesian hierarchical method.}
\label{F:plot1}
\end{center}
\end{figure}

The procedure was repeated to obtain 100 simulated datasets each of size $n=200$.
For a typical dataset, the top panel of Figure \ref{F:plot1} displays the data points using black dots and the true  ellipse using the red solid line. Since the density for the angular parameters has the mode at 0 radians, the data points are more highly concentrated near the left corner of the ellipse upon applying the rotation of $\varphi=\pi$.

\begin{table}
\begin{center}
\renewcommand{\arraystretch}{1}
\begin{tabular}{ l  | c |c  | c}
\hline\hline
\multicolumn{4}{c}{\textbf{\large{Simulation 1}}}\\
\hline
 &\textbf{Long axis}	&\textbf{Short axis}	&\textbf{Center} 	\\
True values &($2a=100$)	&($2b=50$)	&$(c_1,c_2)=(250,250)$ 	 \\
\hline\hline
\multicolumn{4}{c}{\textbf{Biases}}\\
\hline
\textbf{Pseudo inverse} &-0.718     &0.192  &(249.733,  250.019) \\
\textbf{Orthogonal distance}	&-0.194     &0.016  &(249.778,  250.024) \\
\textbf{Bayes}	&-0.420    &-0.024  &(249.780,  250.034) \\
\hline
\multicolumn{4}{c}{\textbf{Standard errors}}\\
\hline
\textbf{Pseudo inverse} &0.142      &0.050    &(0.062,    0.024) \\
\textbf{Orthogonal distance}	&0.140      &0.048    &(0.061,    0.023) \\
\textbf{Bayes}	    	&0.136      &0.046    &(0.059,    0.022) \\
\hline\hline
\end{tabular}
\end{center}
\caption{For Simulation 1, in which data points were generated over the entire ellipse, biases and standard errors for the ellipse parameters averaged over the 100 datasets. The rotational angle is given in radians. See the text for further explanation.}\label{T:simulation1}
\end{table}

Each dataset was analyzed using the methods pseudo inverse \citep[Sec.\ 4.1]{Zhang_1997}, 
orthogonal distance fitting, and the Bayesian approach of Section \ref{Bayesian} conditional on the conic section being an ellipse. The first two methods were also investigated in Section 10.1 of \cite{Zhang_1997}.
For analyzing the angular parameters in the Bayesian approach, following the recommendation of \cite{Turnbull_Ghosh_2014}, we set the degree of the Bernstein polynomial, $m=$$\lceil n / \log n\rceil=$ $38$.

The average bias in the parameter estimates and Monte Carlo (MC) standard errors for the 100 datasets are presented in Table~\ref{T:simulation1}. The  root mean square errors (RMSEs) are then readily available as the sum of the corresponding squared biases and squared  standard errors.  We find that all  three methods provide similar estimates that coincide with the truth within the margin of error.  Averaging over the 100 datasets, the estimated ellipse by the Bayesian hierarchical approach is shown using the blue dashed line in the top panel of
Figure \ref{F:plot1}.

The bottom panel of Figure \ref{F:plot1} displays the true ellipse (black) and the 100 estimated ellipses (gray) by the Bayesian method. We find that the contours of  the estimated ellipses are highly concentrated around the true ellipse; a little more so near the left edge of the ellipse, where more data are available.

\bigskip

\subsection{Simulation Study 2}\label{S:simulation_1}

We  investigated the ability of the Bayesian  procedure to detect the type of conic section from noisy data generated by conics in non-standard form, i.e., hyperbolas, ellipses, and parabolas with unknown focuses and arbitrarily rotated  axes of symmetry. Furthermore, to increase the difficulty of making  calls from a visual inspection, all the data were generated from partial conics, such as half-ellipses that  may be incorrectly detected as hyperbolas or parabolas. Additional variability was introduced into the simulation by randomly generating the true conics parameters for each artificial dataset.

Specifically, we generated 300 artificial datasets with $n=100$ data points each. An equal number of datasets was allocated to true ellipses, parabolas, and hyperbolas. For each dataset, we performed the following steps to randomly generate the conic section parameters and the associated noisy data points:

\begin{enumerate}
\item Referring back to the standard conics equations of Section \ref{S:standard}, the true parameters $a$ and $b$ were generated from normal distributions with means 50 and 25 respectively, and with variances 4. For parabolas, only  parameter $a$ was generated.

\item Rotational angle $\varphi$ was uniformly generated from the interval $(-\pi,\pi)$.

\item The conic center was fixed at $(c_1,c_2)=(250,250)$.

\item The angular parameters for the $n=100$ conic points, defined in equation (\ref{e_support}), were generated as follows:

\begin{enumerate}
\item For ellipses, only one half of the curve was used to generate the data. Specifically, we generated $t_i \stackrel{iid}\sim U(-\pi/2,\pi/2)$.
    \item For hyperbolas, we first computed the eccentricity $e=\sqrt{1+b^2/a^2}$ and  uniformly generated the parameters $t_i$'s over the  range specified in equation~(\ref{e_support}).
        \item For parabolas, we generated the angular parameters as $t_i \stackrel{iid}\sim U(-2,2)$.
    \end{enumerate}

\item Depending on the type of conic section, the semi-latus rectum $l$ and eccentricity $e$ were computed using well-known formulas that involve parameters $a$, $b$ and conic center $(c_1,c_2)$.

    \item Apply equations (\ref{r_i}) and (\ref{xy_i}) to compute $r_i$ and the true conic point, $(u_i,v_i)$.
        \item To both components of the conic points, add independent, zero-mean Gaussian errors with standard deviation $\sigma=2$  to obtain the data, $(x_i,y_i)$ for $i=1,\ldots,n$.
\end{enumerate}

For example, a randomly chosen ellipse, hyperbola, and parabola are  shown in Figure~\ref{F:plot0}. Observe that, being randomly generated, the center,  rotational angle, and other parameters are all different in the three conics. Gaussian random noise with $\sigma=2$ was added to generate each data point in all three conics. However, parabolas and hyperbolas have unbounded  one-dimensional ``surfaces'' or line integrals, and their data have much greater ranges than those of ellipses, as seen in Figure~\ref{F:plot0}. 

Assuming all conic parameters to be unknown, the MCMC procedure based on the categorical variable $\mathcal{C}$ defined in equation (\ref{detectingConic}) was used to identify the type of conic  in each dataset. The type of conic section was correctly detected in 276  (i.e., 92\%) of the datasets. Table \ref{T:simulation01} provides more detail about the estimated posterior probabilities of detection and misclassification for each type of  conic, revealing that ellipses and hyperbolas were rarely or never misclassified. Parabolas were misclassified 21.4\% of the time. A possible explanation is that parabolas are represented by the singleton value of $e=1$, with values to the left representing ellipses and values to the right representing hyperbolas. Averaging over the 300 datasets, the root mean squared error for the eccentricity parameter $e$ was 0.051.

For analyzing these  data, the proposed  technique took, on average, 0.01621 seconds per MCMC iteration on a PC laptop with Intel Core i7 CPU and 8 GB RAM. Despite the high dimensionality of the parameter space,  a few thousand MCMC iterations were found to be more than sufficient  for accurate  posterior inferences due to the fast-mixing MCMC chain. 

The proposed method, like  Bayesian hierarchical approaches in general, is  computationally intensive when compared with several  frequentist approaches for conic sections. However, the benefits  far outweigh the computational costs, since the Bayesian method  not only provides accurate estimates, along with uncertainty estimates, for all parameters of interest, but  also detects  the underlying types of conic section.

\begin{table}[ht!]
\begin{center}
\renewcommand{\arraystretch}{1}
\begin{tabular}{ l  | c |c  | c}
\hline\hline
&\multicolumn{3}{c}{\textit{Detected}}\\
\hline
\textit{True} &\textbf{Ellipse}	&\textbf{Parabola}	&\textbf{Hyperbola} 	\\
\hline
\textbf{Ellipse} &1.000 (0.000) &0.000  (0.000) &0.000  (0.000) \\\hline
\textbf{Parabola}	&0.131  (0.034) &0.786 (0.041) &0.083 (0.027) \\\hline
\textbf{Hyperbola}	&0.000 (0.000) &0.030 (0.017) &0.970 (0.017)\\\hline
\hline
\end{tabular}
\end{center}
\caption{For Simulation 2, Monte Carlo estimates of the posterior probabilities of classification for the three true conic section types. 
The estimated standard errors are displayed in parentheses.}\label{T:simulation01}
\end{table}

\section{Discussion}\label{S:discussion}

We have proposed  a coherent and flexible Bayesian hierarchical model for  fitting all possible conic sections to noisy observations. The methodology is able to accurately compute estimates and uncertainty estimates for all the conic parameters. The success of the technique was demonstrated via simulation studies in which the data were generated from partial as well as complete true ellipses.  In the simulation study, we explored the probabilistic detection of the underlying conics by generating the data from several partial conics, such as half-ellipses that  may be incorrectly detected as hyperbolas or parabolas.  For each artificial dataset in this study, additional variability was introduced by randomly generating the true conics parameters including the rotational angle and conic~type. The simulation results indicate that our  proposed method     has low misclassification rates irrespective of the underlying conic type.

Through further simulation studies, we have found that when the data are generated from only a part of the conic section, our method may produced biased estimates for the conic parameters, even though it is able to correctly identify  the type of conic section. This is also true for the estimates produced by other competing methods in the literature (when they are supplied  addition information about true underlying conic type). Further details of the comparisons can be found in Section 1 of the Appendix.

There is empirical evidence that the proposed technique may lead to large sample consistent estimates. That is, as the number of data points grows, the inference procedure is able to detect the conic section parameters with greater accuracy (e.g., see \cite{Kukush_etal_2004} for consistency using least squares-type methods). In this paper, we have not pursued  theoretical properties such as the posterior consistency of the estimates  and their rates \cite{Ghosal_etal_2017}. Extensions of our methodology to other curved planes and surfaces are also of interest. This would perhaps require a more substantial development than those presented in this paper, and hence remains a part of future developments in this topic.

\newpage
\begin{center}
{\large\bf SUPPLEMENTARY MATERIALS}
\end{center}

\begin{description}


\item[{\tt BayesConics.zip}:] R code for reproducing the  Simulation Study 2 results of Section \ref{S:simulation_1}. The code performs three functions: (a) generates artificial datasets, (b) analyzes the generated data using the proposed Bayesian methodology to detect the underlying conic type, and (c) provides a summary of the detection accuracy, presented in Table 3. The README file in the zip folder contains instructions for use.

\end{description}

\bibliographystyle{Perfect}

\bibliography{main}
\end{document}